\documentclass[pra,10pt,aps,twocolumn,amsmath,amssymb,superscriptaddress,altaffilletter]{revtex4-1}

\usepackage     [utf8]                  {inputenc}
\usepackage     [T1]                    {fontenc}
\usepackage                             {color}
\usepackage                             {graphicx}
\usepackage     [english]               {babel}
\usepackage     [colorlinks,citecolor=black,linkcolor=black,urlcolor=blue,bookmarks=false,hypertexnames=true]                        {hyperref}
\usepackage	[usenames,dvipsnames]	{pstricks}
\usepackage				{dcolumn}
\usepackage				{bm}
\usepackage				{textcomp}

\begin{document}

\title{Thermalization as an Invisibility Cloak for Fragile Quantum Superpositions}

\author{Walter Hahn}
\email{w.hahn@skoltech.ru}
\affiliation{Skolkovo Institute of Science and Technology, Skolkovo Innovation Centre, Nobel Street 3, Moscow 143026, Russia}

\author{Boris V. Fine}
\email{b.fine@skoltech.ru}
\affiliation{Skolkovo Institute of Science and Technology, Skolkovo Innovation Centre, Nobel Street 3, Moscow 143026, Russia}
\affiliation{Institute for Theoretical Physics, Philosophenweg 12, 69120 Heidelberg, Germany}

\begin{abstract}
We propose a method for protecting fragile quantum superpositions in many-particle systems from dephasing by external classical noise. We call superpositions ``fragile'' if dephasing occurs particularly fast, because the noise couples very differently to the superposed states. The method consists of letting a quantum superposition evolve under the internal thermalization dynamics of the system, followed by a time reversal manipulation known as Loschmidt echo. The thermalization dynamics makes the superposed states almost indistinguishable during most of the above procedure. We validate the method by applying it to a cluster of spins-\textonehalf.
\end{abstract}


\maketitle

The ability of quantum systems to exhibit interference between different quantum states is central to various fields of modern-day research, ranging from quantum simulator technology to the foundations of quantum theory. In the latter case, observing interference involving an increasingly large number of particles serves for testing the applicability limits of quantum mechanics~\cite{leggett,hornberger_arndt}. Quantum interference requires coherent quantum superpositions. However, quantum coherence is typically lost as a result of interactions between the system of interest and the environment. This loss of coherence can be categorized into two processes~\cite{joos_zeh,zurek}: either the system becomes entangled with the environment and, therefore, it is no longer in a pure quantum state, or the environment acts as a classical time-dependent noise inducing dephasing. In this work, we focus on the latter and propose a method for protecting quantum superpositions from dephasing for generic many-body quantum systems. Our primary interest is in quantum superpositions which are particularly susceptible to dephasing, because the noise couples to an extensive variable characterizing the system and, at the same time, the superposed states have very different expectation values of that variable. We call such superpositions ``fragile''. For example, for a quantum spin cluster in a fluctuating magnetic field, the superposition of states with all spins ``up'' or ``down'' along any axis, \mbox{$[|\!\uparrow\uparrow\uparrow\cdots\rangle+|\!\downarrow\downarrow\downarrow\cdots\rangle]/\sqrt{2}$, is fragile}. The notion of fragility is related to the notion of macroscopic quantum superpositions - see, for example, Refs.~\cite{macr,macr_fr}.

The idea of the method is to use the thermalization dynamics within a many-body system as an invisibility cloak for coherent superpositions. We assume that, in the system of interest, the internal interactions can be controlled experimentally. In such a case, it would seem natural, at first sight, to protect a coherent superposition by switching off the interactions completely, because they are known to cause internal decoherence on top of dephasing due to the external noise. However, as we show below, there is a better alternative, namely, to let a quantum superposition initially evolve under the internal dynamics of the system, then to reverse this dynamics at time $\tau_0$ by changing the sign of the interaction Hamiltonian, and, finally, to recover the initial superposition at time $2\tau_0$. Such a procedure is used to generate the so-called ``Loschmidt echo''~\cite{pastawski_2,prosen_loschmidt,goussev}, also known as ``magic echo'' in nuclear magnetic resonance~\cite{hahn_echo,pines_waugh,slichter}. A Loschmidt-echo manipulation in the presence of internal interactions not only reverses internal decoherence but also suppresses dephasing due to external noise. The interaction Hamiltonian of the system must be chosen such that, for each of the superposed states, the expectation value of the variable coupled to the noise decays on the timescale much faster than $\tau_0$ and, as a result, the superposed states become much less distinguishable for the noise during most of the time interval $[0,2\tau_0]$. In addition, after the decay, the above variable exhibits fast fluctuations caused by internal dynamics, which further reduces the effective coupling to the noise, thereby also suppressing dephasing. 

The above method for protecting quantum superpositions is complementary to existing methods~\cite{suter_rmp} which use dynamical decoupling~\cite{viola_lloyd}, decoherence-free subspaces~\cite{wineland_memory,viola_noiseless}, feedback schemes~\cite{milburn_feedback,control_2}, and quantum-memory techniques~\cite{haroche_memory,reversible_1}. Our method is particularly suitable for dealing with large quantum systems in situations where both external noise and the internal decoherence are present and the internal decoherence cannot be reversed by the conventional Hahn-echo technique~\cite{hahn_echo}.

In the following, we illustrate the above method by applying it to lattices of $N_s$ spins-\textonehalf. Such a lattice can, for example, represent a cluster of nuclear spins in a solid. We start with a general description of dephasing for a noninteracting spin system in a fluctuating magnetic field. Then, we extend this description by adding interactions between spins together with the Loschmidt-echo sequence. Finally, we compare the two cases by means of direct calculations.

Let us consider Hamiltonian ${\cal H}_\text{N}=h(t)\sum_j{S_{jz}}$, where $S_{jz}$ is the spin-\textonehalf\ $z$-projection operator for the $j$th lattice site, and $h(t)$ is a fluctuating magnetic field along the $z$ axis. The field is characterized by the time correlation function $\langle h(0)h(t)\rangle=h_\text{rms}^2\exp(-\gamma t)$, where $h_\text{rms}$ is the root-mean-squared value of $h(t)$ and $\gamma$ is the decay constant. We consider $h_\text{rms}\ll\gamma$. In our calculations, we use $h_\text{rms}=0.0085$, $\gamma=\frac{1}{4}$, and $\hbar=1$.

Let us further assume that the superposition that we want to protect has the form
\begin{equation} \label{eqn_init}
 |\Psi(0)\rangle=\frac{1}{\sqrt{2}}\Big[|\psi_1\rangle+|\psi_2\rangle\Big],
\end{equation}
where $|\psi_1\rangle$ and $|\psi_2\rangle$ are eigenstates of the total-magnetization operator $M_z\equiv\sum_j{S_{jz}}$ corresponding to the magnetization values $M_{z,1}$ and $M_{z,2}$, respectively. We choose $M_{z,1}$ and $M_{z,2}$ such that $|M_{z,2}-M_{z,1}|\sim N_s$. Since $M_z$ is the variable that the noise $h(t)$ couples to, the large value of $|M_{z,2}-M_{z,1}|$ implies that, according to our definition, $|\Psi(0)\rangle$ is a fragile superposition.

\begin{figure}[b]
	\centering
	\includegraphics[width=0.9\columnwidth]{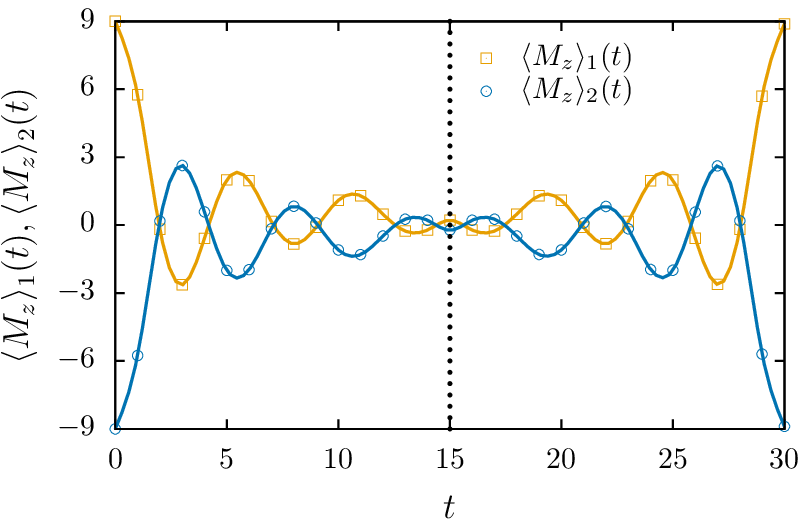}
	\caption{Magnetizations $\langle M_z\rangle_1(t)$ and $\langle M_z\rangle_2(t)$ for initial quantum states \mbox{$|\psi_1\rangle=|\!\uparrow\uparrow\uparrow\cdots\rangle$} and \mbox{$|\psi_2\rangle=|\!\downarrow\downarrow\downarrow\cdots\rangle$} of a periodic chain of 18 spins-\textonehalf\ governed by Hamiltonian~\eqref{eqn_ham}. Symbols represent numerical calculations, lines are guides to the eye. The interaction Hamiltonian is reversed at $\tau_0=15$ indicated by the dotted line. This reversal leads to a nearly perfect revival at $t=2\tau_0$.}
	\label{fig_magnetization}
\end{figure}

Starting with $|\Psi(0)\rangle$ given by Eq.~\eqref{eqn_init}, we obtain at later times
\begin{equation} \label{eqn_wf}
 |\Psi(t)\rangle=\frac{1}{\sqrt{2}}\Big[e^{-i\varphi_1(t)}|\psi_1\rangle+e^{-i\varphi_2(t)}|\psi_2\rangle\Big],
\end{equation}
where the acquired phase difference is
\begin{equation} \label{eqn_phase}
 \Delta\varphi(t)\equiv\varphi_2(t)-\varphi_1(t)=(M_{z,2}-M_{z,1})\int_0^th(t')dt'.
\end{equation}
Since $h(t)$ randomly fluctuates, $\Delta\varphi(t)$ exhibits a diffusive random behavior. While coherence is preserved by unitary dynamics in each individual realization of $h(t)$, the ensemble describing all possible realizations of $h(t)$ exhibits a coherence decay.

Now let us assume that the spin cluster considered is a periodic chain with nearest-neighbor (NN) interaction, where we can engineer interactions, such that the Hamiltonian becomes
\begin{equation} \label{eqn_ham}
 {\cal H}_\text{I}=\!\sum^{NN}_{i,j}\Big[J_x S_{ix}S_{jx}+J_y S_{iy}S_{jy}+J_z S_{iz}S_{jz}\Big]+h(t)\sum_j{S_{jz}},
\end{equation}
where $J_x$, $J_y$, and $J_z$ are interaction constants satisfying \mbox{$J_\text{eff}\gg\gamma$} with \mbox{$J_\text{eff}^2\equiv J_x^2+J_y^2+J_z^2$}. The last term in Eq.~\eqref{eqn_ham} equals ${\cal H}_\text{N}$ defined earlier. The Loschmidt echo is to be implemented by changing the sign of all interaction constants $\{J_x,J_y,J_z\}\rightarrow\{-J_x,-J_y,-J_z\}$ at time $t=\tau_0$. Below, we specifically consider $J_x=-0.47$, $J_y=0.79$, $J_z=0.37$, and \mbox{$|\psi_1\rangle=|\!\uparrow\uparrow\uparrow\cdots\rangle$}, \mbox{$|\psi_2\rangle=|\!\downarrow\downarrow\downarrow\cdots\rangle$}, where \mbox{$|\!\uparrow\rangle$} and \mbox{$|\!\downarrow\rangle$} describe spins pointing up and down along the $z$ axis. The corresponding values of magnetization are \mbox{$M_{z,1}=-M_{z,2}=\frac{N_s}{2}$}.

The values of the interaction constants are chosen such that magnetization $M_z$ is not conserved, i.e., $[{\cal H}_\text{I},M_z]\neq0$, and, moreover, for large $t$ (but still smaller than $\tau_0$)
\begin{equation} \label{eqn_decay}
 \langle M_z\rangle_1(t)-\langle M_z\rangle_2(t)\rightarrow 0,
\end{equation}
where $\langle M_z\rangle_1(t)\!\equiv\!\langle\psi_1|M_z(t)|\psi_1\rangle$, $\langle M_z\rangle_2(t)\!\equiv\!\langle\psi_2|M_z(t)|\psi_2\rangle$. [Here, $M_z(t)$ is an operator in the Heisenberg representation.] After the decay of $\langle M_z\rangle_1(t)$ and $\langle M_z\rangle_2(t)$ shown in Fig.~\ref{fig_magnetization}, the noise can distinguish between the superposed states only on the basis of fluctuations of the order $\sqrt{N_s}$, whereas, initially, the superposed states were distinguishable by their total magnetization of the order $N_s$ (as in the noninteracting case).

\begin{figure}[t]
	\centering
	\includegraphics[width=0.9\columnwidth]{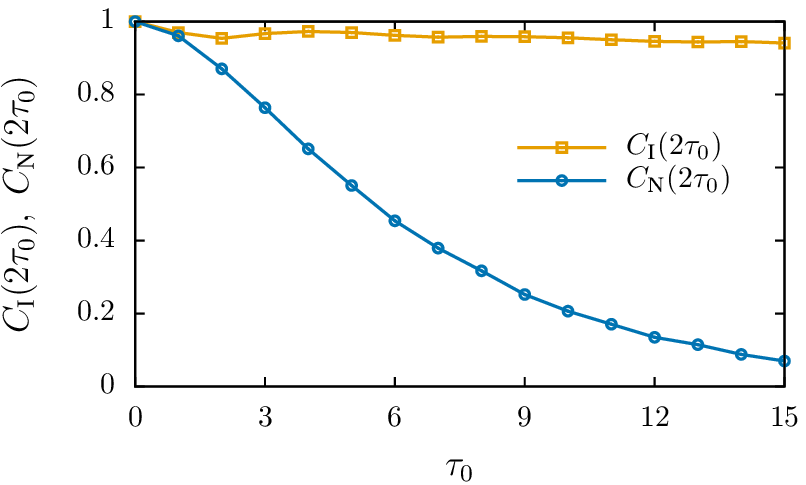}
	\caption{Coherence measures $C_\text{I}(2\tau_0)$ and $C_\text{N}(2\tau_0)$ computed respectively for interacting and noninteracting periodic chains of 18 spins-\textonehalf\ in the presence of external noise. Symbols represent numerical calculations, lines are guides to the eye.}
	\label{fig_example}
\end{figure}

\begin{figure*}[t]
	\centering
	\includegraphics[width=0.95\textwidth]{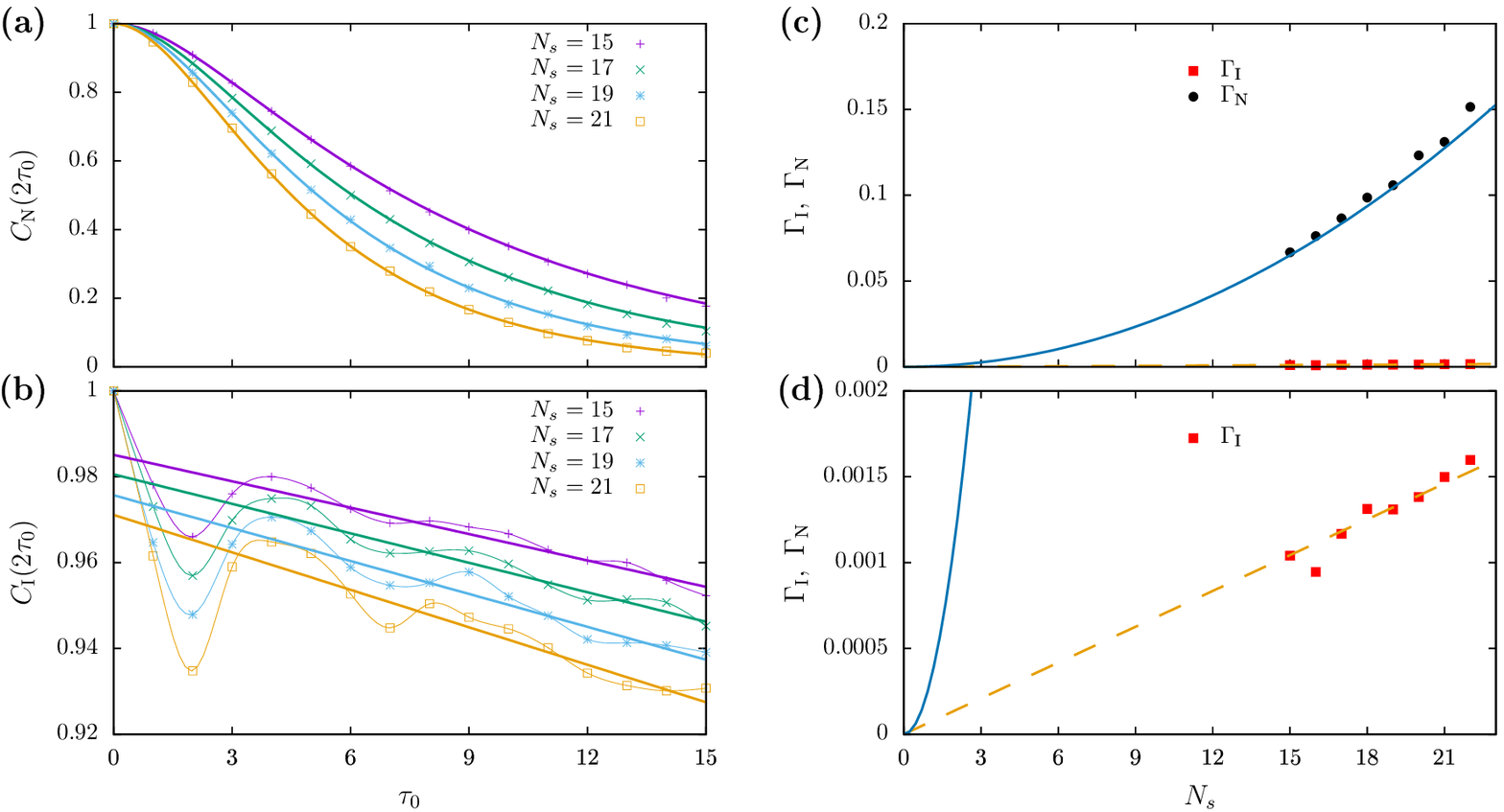}
	\caption{Decay of coherence measures as a function of the number of spins $N_s$: (a) noninteracting case, (b) interacting case. In (a) and (b), symbols represent numerically computed $C_\text{N}(2\tau_0)$ and $C_\text{I}(2\tau_0)$, respectively, corresponding to $N_s$ indicated in plot legends; thin lines connecting symbols are guides to the eye; thick lines represent Eq.~\eqref{eqn_and_weiss} for (a) and linear function $C_\text{I}(2\tau_0)\approx b(1-2\Gamma_\text{I}\tau_0)$ with fitted parameters $b$ and $\Gamma_\text{I}$ for (b). [Note the different scale of the vertical axis in (a) and (b).] (c), (d) Asymptotic decay rates $\Gamma_\text{N}$ and $\Gamma_\text{I}$ for the noninteracting and interacting cases, respectively, as a function of $N_s$. Symbols represent the values obtained by fitting the tails of $C_\text{N}(2\tau_0)$ and $C_\text{I}(2\tau_0)$; solid blue line represents Eq.~\eqref{eqn_gamman}; dashed orange line represents Eq.~\eqref{eqn_gammai} with a fitted prefactor 0.96. Plots (c) and (d) differ only by the scale of the vertical axis.}
	\label{fig_cc}
\end{figure*}

The reversal of the Hamiltonian ${\cal H}_\text{I}$ at $t=\tau_0$ is only partial, because it does not apply to the noise term. Therefore, the wave function $|\Psi(2\tau_0)\rangle$ is not expected to coincide with $|\Psi(0)\rangle$. Instead, we parametrize it as
\begin{equation} \label{eqn_wf2}
 |\Psi(2\tau_0)\rangle=c_1(2\tau_0)|\psi_1\rangle+c_2(2\tau_0)|\psi_2\rangle+c_\phi(2\tau_0)|\phi(2\tau_0)\rangle,
\end{equation}
where $|\phi(2\tau_0)\rangle$ is a state orthogonal to $|\psi_1\rangle$ and $|\psi_2\rangle$, and $c_1(2\tau_0)$, $c_2(2\tau_0)$, and $c_\phi(2\tau_0)$ are complex amplitudes. The same parametrization is also applicable to the noninteracting case described by Eq.~\eqref{eqn_wf} [with $c_\phi(2\tau_0)=0$].

To compare the coherence loss at time $2\tau_0$ for the interacting and noninteracting cases, we introduce the coherence measure
\begin{equation} \label{eqn_ct}
 C(2\tau_0)\equiv2\big|\big\langle c_1^*(2\tau_0)c_2(2\tau_0)\big\rangle\big|,
\end{equation}
where the angle brackets denote averaging over all possible realizations of $h(t)$. The term $c_1^*(2\tau_0)c_2(2\tau_0)$ is the off-diagonal element of system's density matrix connecting the states $|\psi_1\rangle$ and $|\psi_2\rangle$. The coherence measure $C(2\tau_0)$ is related to the quantum fidelity (see, e.g., Ref.~\cite{coh}). The value $C(2\tau_0)=1$ implies that the initial coherence between $|\psi_1\rangle$ and $|\psi_2\rangle$ is fully retained, while $C(2\tau_0)=0$ means that it is completely lost. Below, functions $C_\text{N}(2\tau_0)$ and $C_\text{I}(2\tau_0)$ represent $C(2\tau_0)$ computed for the noninteracting (${\cal H}_\text{N}$) and interacting (${\cal H}_\text{I}$) cases, respectively.

In simulations, we represent the noise by a sufficiently dense set of discrete Fourier harmonics \mbox{$h(t)=\sum_j h_{\omega_j}\cos(\omega_jt+\alpha_j)$}, where $\omega_j$ are frequencies~\footnote{We choose $\omega_j=\Delta_\omega j$ with $\Delta_\omega=\pi/1000$, and $j$ is an integer $j\in[-100000,100000]$}, $\alpha_j$ random phases, and \mbox{$h_{\omega_j}=Ah_\text{rms}/\sqrt{\omega_j^2+\gamma^2}$}, with $A$ being a normalization constant.

In the noninteracting case, we substitute the complex amplitudes $c_1(2\tau_0)$ and $c_2(2\tau_0)$ from Eq.~\eqref{eqn_wf} into definition~\eqref{eqn_ct}, thereby obtaining \mbox{$C_\text{N}(2\tau_0)=|\langle \exp\{i[\varphi_1(2\tau_0)-\varphi_2(2\tau_0)]\}\rangle|$}. We then use Eq.~\eqref{eqn_phase}, where we perform explicit time integration to obtain
\begin{eqnarray} \label{eqn_aver}
 &&C_\text{N}(2\tau_0)=\\
 &&\!\left|\!\left\langle\!\!\cos\!\!\left[\!2(M_{z,2}-M_{z,1})\!\sum_j h_{\omega_j}\frac{\sin(\omega_j\tau_0)}{\omega_j}\!\cos\!\left(\omega_j\tau_0\!+\!\alpha_j\right)\!\right]\!\right\rangle\!\right|. \nonumber
\end{eqnarray}
Finally, we calculate $C_\text{N}(2\tau_0)$ by averaging of the above expression numerically over completely random phases $\alpha_j$.

For the interacting case, we calculate $C_\text{I}(2\tau_0)$ numerically by means of direct integration of the Schrödinger equation~\cite{tarek,steinigeweg}, which does not require complete diagonalization of the Hamiltonian. When implemented with a fourth-order Runge-Kutta algorithm, the above method is shown~\cite{tarek} to be very accurate for the time intervals of interest.

Typical behavior of $C_\text{N}(2\tau_0)$ and $C_\text{I}(2\tau_0)$ is shown in Fig.~\ref{fig_example}. Initially, for $\tau_0\lesssim1$, $C_\text{N}(2\tau_0)\approx C_\text{I}(2\tau_0)$. At later times, $C_\text{I}(2\tau_0)$ decays much slower than $C_\text{N}(2\tau_0)$ which demonstrates the effectiveness of the proposed method.

Let us now investigate the decay of $C_\text{N}(2\tau_0)$ and $C_\text{I}(2\tau_0)$ for different system sizes $N_s$.

For the noninteracting case, the results of our simulations based on Eq.~\eqref{eqn_aver} are shown in Fig.~\ref{fig_cc} (a). These results are in excellent agreement with the theoretical approximation of Anderson and Weiss~\cite{and_weiss}
\begin{equation} \label{eqn_and_weiss}
 C_\text{N}(2\tau_0)\approx\exp\left[-\omega_\varphi^2\int_0^{2\tau_0}(2\tau_0-t')\Delta(t')dt'\right],
\end{equation}
where $\Delta(t')\equiv \langle\Delta\varphi(0)\Delta\varphi(t')\rangle/\langle\Delta\varphi^2(0)\rangle$ and $\omega_\varphi^2\equiv \langle\Delta\varphi^2(0)\rangle$, cf. Eq.~\eqref{eqn_phase}. In our case, $\omega_\varphi^2=N_s^2h_\text{rms}^2$ and
$\Delta(t')=e^{-\gamma t'}$. Function $C_\text{N}(2\tau_0)$ given by Eq.~\eqref{eqn_and_weiss} starts decaying as a Gaussian and then approaches the asymptotic exponential regime characterized by the decay rate
\begin{equation} \label{eqn_gamman}
 \Gamma_\text{N}=N_s^2\frac{h_\text{rms}^2}{\gamma}.
\end{equation}
In Fig.~\ref{fig_cc} (c), we plot the values of $\Gamma_\text{N}$ obtained numerically as a function of $N_s$ together with the right-hand side of Eq.~\eqref{eqn_gamman}.

Numerically computed functions $C_\text{I}(2\tau_0)$ are shown for different $N_s$ in \mbox{Fig.~\ref{fig_cc} (b)}. The initial oscillations of $C_\text{I}(2\tau_0)$ are presumably due to the oscillations of the magnetization shown in Fig.~\ref{fig_magnetization}. We expect the overall decay of $C_\text{I}(2\tau_0)$ on a longer timescale to be exponential. However, on time scales accessible numerically, we only observe the initial decay and, therefore, fit $C_\text{I}(2\tau_0)$ with a linear function $C_\text{I}(2\tau_0)\approx b(1-2\Gamma_\text{I}\tau_0)$, where $b$ is a prefactor and $\Gamma_\text{I}$ is the characteristic exponential decay rate plotted in Figs.~\ref{fig_cc} (c) and \ref{fig_cc} (d). On the basis of analysis similar to that in Ref.~\cite{and_weiss}, we estimate
\begin{equation} \label{eqn_gammai}
 \Gamma_\text{I}\sim N_s\frac{h_\text{rms}^2}{J_\text{eff}},
\end{equation}
which is consistent with our numerical results, as illustrated in Fig.~\ref{fig_cc} (d).

Given Eqs.~\eqref{eqn_gamman} and~\eqref{eqn_gammai}, we obtain
\begin{equation} \label{eqn_relation}
 \frac{\Gamma_\text{N}}{\Gamma_\text{I}}\sim N_s\frac{J_\text{eff}}{\gamma}.
\end{equation}
The larger this ratio, the more effective our method. Substituting typical parameters of our simulations into Eq.~\eqref{eqn_relation}, we obtain $\Gamma_\text{N}/\Gamma_\text{I}\sim10^2$. In general, Eq.~\eqref{eqn_relation} implies that the proposed method becomes more effective when the system becomes larger and its internal dynamics becomes faster. Since the method only requires the system of interest to thermalize much faster than the correlation time of the noise, we expect it to be applicable to a broad class of fragile quantum superpositions in a broad class of quantum systems, where Loschmidt-echo is experimentally realizable~\cite{pines_waugh,negative_temp,andrey}.

Let us now make two final remarks:

(i) A necessary requirement for the proposed method is a sufficiently accurate reversal of the interaction Hamiltonian. We estimate the acceptable deviation of the experimentally reversed Hamiltonian from the perfectly reversed one to be (per spin) of the order of $h_\text{rms}$ or smaller~\cite{absence}.

(ii) The proposed method can also be used for protecting fragile quantum superpositions from decoherence by external particles which are sufficiently slow and/or couple simultaneously to sufficiently many particles within the system. A relevant example here is the decoherence by long-wavelength photons.

To conclude, we have proposed a Loschmidt-echo based method for protecting fragile quantum superpositions in many-body systems and validated this method by both numerical simulations and analytical estimates. In the concrete examples considered, the lifetime of coherent superpositions was shown to increase by two orders of magnitude.

\textit{Acknowledgements} - W.H. is grateful for the hospitality of the \textit{Massachusetts Institute of Technology}, where a part of this work was done. This work was supported by grant of \textit{Russian Science Foundation} (project number 17-12-01587).


\begin{thebibliography}{28}%
\makeatletter
\providecommand \@ifxundefined [1]{%
 \@ifx{#1\undefined}
}%
\providecommand \@ifnum [1]{%
 \ifnum #1\expandafter \@firstoftwo
 \else \expandafter \@secondoftwo
 \fi
}%
\providecommand \@ifx [1]{%
 \ifx #1\expandafter \@firstoftwo
 \else \expandafter \@secondoftwo
 \fi
}%
\providecommand \natexlab [1]{#1}%
\providecommand \enquote  [1]{``#1''}%
\providecommand \bibnamefont  [1]{#1}%
\providecommand \bibfnamefont [1]{#1}%
\providecommand \citenamefont [1]{#1}%
\providecommand \href@noop [0]{\@secondoftwo}%
\providecommand \href [0]{\begingroup \@sanitize@url \@href}%
\providecommand \@href[1]{\@@startlink{#1}\@@href}%
\providecommand \@@href[1]{\endgroup#1\@@endlink}%
\providecommand \@sanitize@url [0]{\catcode `\\12\catcode `\$12\catcode
  `\&12\catcode `\#12\catcode `\^12\catcode `\_12\catcode `\%12\relax}%
\providecommand \@@startlink[1]{}%
\providecommand \@@endlink[0]{}%
\providecommand \url  [0]{\begingroup\@sanitize@url \@url }%
\providecommand \@url [1]{\endgroup\@href {#1}{\urlprefix }}%
\providecommand \urlprefix  [0]{URL }%
\providecommand \Eprint [0]{\href }%
\providecommand \doibase [0]{http://dx.doi.org/}%
\providecommand \selectlanguage [0]{\@gobble}%
\providecommand \bibinfo  [0]{\@secondoftwo}%
\providecommand \bibfield  [0]{\@secondoftwo}%
\providecommand \translation [1]{[#1]}%
\providecommand \BibitemOpen [0]{}%
\providecommand \bibitemStop [0]{}%
\providecommand \bibitemNoStop [0]{.\EOS\space}%
\providecommand \EOS [0]{\spacefactor3000\relax}%
\providecommand \BibitemShut  [1]{\csname bibitem#1\endcsname}%
\let\auto@bib@innerbib\@empty
\bibitem [{\citenamefont {Leggett}(2002)}]{leggett}%
  \BibitemOpen
  \bibfield  {author} {\bibinfo {author} {\bibfnamefont {A.~J.}\ \bibnamefont
  {Leggett}},\ }\href {http://stacks.iop.org/0953-8984/14/i=15/a=201}
  {\bibfield  {journal} {\bibinfo  {journal} {Journal of Physics: Condensed
  Matter}\ }\textbf {\bibinfo {volume} {14}},\ \bibinfo {pages} {R415}
  (\bibinfo {year} {2002})}\BibitemShut {NoStop}%
\bibitem [{\citenamefont {Arndt}\ and\ \citenamefont
  {Hornberger}(2014)}]{hornberger_arndt}%
  \BibitemOpen
  \bibfield  {author} {\bibinfo {author} {\bibfnamefont {M.}~\bibnamefont
  {Arndt}}\ and\ \bibinfo {author} {\bibfnamefont {K.}~\bibnamefont
  {Hornberger}},\ }\href {\doibase 10.1038/nphys2863} {\bibfield  {journal}
  {\bibinfo  {journal} {Nature Physics}\ }\textbf {\bibinfo {volume} {10}},\
  \bibinfo {pages} {271} (\bibinfo {year} {2014})}\BibitemShut {NoStop}%
\bibitem [{\citenamefont {Joos}\ \emph {et~al.}(2003)\citenamefont {Joos},
  \citenamefont {Zeh}, \citenamefont {Kiefer}, \citenamefont {Giulini},
  \citenamefont {Kupsch},\ and\ \citenamefont {Stamatescu}}]{joos_zeh}%
  \BibitemOpen
  \bibfield  {author} {\bibinfo {author} {\bibfnamefont {E.}~\bibnamefont
  {Joos}}, \bibinfo {author} {\bibfnamefont {H.~D.}\ \bibnamefont {Zeh}},
  \bibinfo {author} {\bibfnamefont {C.}~\bibnamefont {Kiefer}}, \bibinfo
  {author} {\bibfnamefont {D.~J.~W.}\ \bibnamefont {Giulini}}, \bibinfo
  {author} {\bibfnamefont {J.}~\bibnamefont {Kupsch}}, \ and\ \bibinfo {author}
  {\bibfnamefont {I.-O.}\ \bibnamefont {Stamatescu}},\ }\href@noop {} {\emph
  {\bibinfo {title} {Decoherence and the Appearance of a Classical World in
  Quantum Theory}}}\ (\bibinfo  {publisher} {Springer-Verlag Berlin
  Heidelberg},\ \bibinfo {year} {2003})\BibitemShut {NoStop}%
\bibitem [{\citenamefont {Zurek}(2003)}]{zurek}%
  \BibitemOpen
  \bibfield  {author} {\bibinfo {author} {\bibfnamefont {W.~H.}\ \bibnamefont
  {Zurek}},\ }\href {\doibase 10.1103/RevModPhys.75.715} {\bibfield  {journal}
  {\bibinfo  {journal} {Rev. Mod. Phys.}\ }\textbf {\bibinfo {volume} {75}},\
  \bibinfo {pages} {715} (\bibinfo {year} {2003})}\BibitemShut {NoStop}%
\bibitem [{\citenamefont {Jeong}\ \emph {et~al.}(2015)\citenamefont {Jeong},
  \citenamefont {Kang},\ and\ \citenamefont {Kwon}}]{macr}%
  \BibitemOpen
  \bibfield  {author} {\bibinfo {author} {\bibfnamefont {H.}~\bibnamefont
  {Jeong}}, \bibinfo {author} {\bibfnamefont {M.}~\bibnamefont {Kang}}, \ and\
  \bibinfo {author} {\bibfnamefont {H.}~\bibnamefont {Kwon}},\ }\href {\doibase
  http://dx.doi.org/10.1016/j.optcom.2014.07.012} {\bibfield  {journal}
  {\bibinfo  {journal} {Optics Communications}\ }\textbf {\bibinfo {volume}
  {337}},\ \bibinfo {pages} {12 } (\bibinfo {year} {2015})}\BibitemShut
  {NoStop}%
\bibitem [{\citenamefont {Fröwis}\ \emph {et~al.}()\citenamefont {Fröwis},
  \citenamefont {Sekatski}, \citenamefont {Dür}, \citenamefont {Gisin},\ and\
  \citenamefont {Sangouard}}]{macr_fr}%
  \BibitemOpen
  \bibfield  {author} {\bibinfo {author} {\bibfnamefont {F.}~\bibnamefont
  {Fröwis}}, \bibinfo {author} {\bibfnamefont {P.}~\bibnamefont {Sekatski}},
  \bibinfo {author} {\bibfnamefont {W.}~\bibnamefont {Dür}}, \bibinfo {author}
  {\bibfnamefont {N.}~\bibnamefont {Gisin}}, \ and\ \bibinfo {author}
  {\bibfnamefont {N.}~\bibnamefont {Sangouard}},\ }\href
  {https://arxiv.org/abs/1706.06173} {\bibinfo  {journal} {arXiv:1706.06173}\
  }\BibitemShut {NoStop}%
\bibitem [{\citenamefont {Pastawski}\ \emph {et~al.}(2000)\citenamefont
  {Pastawski}, \citenamefont {Levstein}, \citenamefont {Usaj}, \citenamefont
  {Raya},\ and\ \citenamefont {Hirschinger}}]{pastawski_2}%
  \BibitemOpen
\bibfield  {journal} {  }\bibfield  {author} {\bibinfo {author} {\bibfnamefont
  {H.}~\bibnamefont {Pastawski}}, \bibinfo {author} {\bibfnamefont
  {P.}~\bibnamefont {Levstein}}, \bibinfo {author} {\bibfnamefont
  {G.}~\bibnamefont {Usaj}}, \bibinfo {author} {\bibfnamefont {J.}~\bibnamefont
  {Raya}}, \ and\ \bibinfo {author} {\bibfnamefont {J.}~\bibnamefont
  {Hirschinger}},\ }\href {\doibase
  http://dx.doi.org/10.1016/S0378-4371(00)00146-1} {\bibfield  {journal}
  {\bibinfo  {journal} {Physica A: Statistical Mechanics and its Applications}\
  }\textbf {\bibinfo {volume} {283}},\ \bibinfo {pages} {166 } (\bibinfo {year}
  {2000})}\BibitemShut {NoStop}%
\bibitem [{\citenamefont {Gorin}\ \emph {et~al.}(2006)\citenamefont {Gorin},
  \citenamefont {Prosen}, \citenamefont {Seligman},\ and\ \citenamefont
  {Žnidarič}}]{prosen_loschmidt}%
  \BibitemOpen
  \bibfield  {author} {\bibinfo {author} {\bibfnamefont {T.}~\bibnamefont
  {Gorin}}, \bibinfo {author} {\bibfnamefont {T.}~\bibnamefont {Prosen}},
  \bibinfo {author} {\bibfnamefont {T.~H.}\ \bibnamefont {Seligman}}, \ and\
  \bibinfo {author} {\bibfnamefont {M.}~\bibnamefont {Žnidarič}},\ }\href
  {\doibase http://dx.doi.org/10.1016/j.physrep.2006.09.003} {\bibfield
  {journal} {\bibinfo  {journal} {Physics Reports}\ }\textbf {\bibinfo {volume}
  {435}},\ \bibinfo {pages} {33 } (\bibinfo {year} {2006})}\BibitemShut
  {NoStop}%
\bibitem [{\citenamefont {Goussev}\ and\ \citenamefont
  {Richter}(2007)}]{goussev}%
  \BibitemOpen
  \bibfield  {author} {\bibinfo {author} {\bibfnamefont {A.}~\bibnamefont
  {Goussev}}\ and\ \bibinfo {author} {\bibfnamefont {K.}~\bibnamefont
  {Richter}},\ }\href {\doibase 10.1103/PhysRevE.75.015201} {\bibfield
  {journal} {\bibinfo  {journal} {Phys. Rev. E}\ }\textbf {\bibinfo {volume}
  {75}},\ \bibinfo {pages} {015201} (\bibinfo {year} {2007})}\BibitemShut
  {NoStop}%
\bibitem [{\citenamefont {Hahn}(1950)}]{hahn_echo}%
  \BibitemOpen
  \bibfield  {author} {\bibinfo {author} {\bibfnamefont {E.~L.}\ \bibnamefont
  {Hahn}},\ }\href {\doibase 10.1103/PhysRev.80.580} {\bibfield  {journal}
  {\bibinfo  {journal} {Phys. Rev.}\ }\textbf {\bibinfo {volume} {80}},\
  \bibinfo {pages} {580} (\bibinfo {year} {1950})}\BibitemShut {NoStop}%
\bibitem [{\citenamefont {Rhim}\ \emph {et~al.}(1971)\citenamefont {Rhim},
  \citenamefont {Pines},\ and\ \citenamefont {Waugh}}]{pines_waugh}%
  \BibitemOpen
  \bibfield  {author} {\bibinfo {author} {\bibfnamefont {W.-K.}\ \bibnamefont
  {Rhim}}, \bibinfo {author} {\bibfnamefont {A.}~\bibnamefont {Pines}}, \ and\
  \bibinfo {author} {\bibfnamefont {J.~S.}\ \bibnamefont {Waugh}},\ }\href
  {\doibase 10.1103/PhysRevB.3.684} {\bibfield  {journal} {\bibinfo  {journal}
  {Phys. Rev. B}\ }\textbf {\bibinfo {volume} {3}},\ \bibinfo {pages} {684}
  (\bibinfo {year} {1971})}\BibitemShut {NoStop}%
\bibitem [{\citenamefont {Slichter}(1990)}]{slichter}%
  \BibitemOpen
  \bibfield  {author} {\bibinfo {author} {\bibfnamefont {C.~P.}\ \bibnamefont
  {Slichter}},\ }\href@noop {} {\emph {\bibinfo {title} {Principles of Magnetic
  Resonance}}}\ (\bibinfo  {publisher} {Springer-Verlag Berlin Heidelberg},\
  \bibinfo {year} {1990})\BibitemShut {NoStop}%
\bibitem [{\citenamefont {Suter}\ and\ \citenamefont
  {\'Alvarez}(2016)}]{suter_rmp}%
  \BibitemOpen
  \bibfield  {author} {\bibinfo {author} {\bibfnamefont {D.}~\bibnamefont
  {Suter}}\ and\ \bibinfo {author} {\bibfnamefont {G.~A.}\ \bibnamefont
  {\'Alvarez}},\ }\href {\doibase 10.1103/RevModPhys.88.041001} {\bibfield
  {journal} {\bibinfo  {journal} {Rev. Mod. Phys.}\ }\textbf {\bibinfo {volume}
  {88}},\ \bibinfo {pages} {041001} (\bibinfo {year} {2016})}\BibitemShut
  {NoStop}%
\bibitem [{\citenamefont {Viola}\ and\ \citenamefont
  {Lloyd}(1998)}]{viola_lloyd}%
  \BibitemOpen
  \bibfield  {author} {\bibinfo {author} {\bibfnamefont {L.}~\bibnamefont
  {Viola}}\ and\ \bibinfo {author} {\bibfnamefont {S.}~\bibnamefont {Lloyd}},\
  }\href {\doibase 10.1103/PhysRevA.58.2733} {\bibfield  {journal} {\bibinfo
  {journal} {Phys. Rev. A}\ }\textbf {\bibinfo {volume} {58}},\ \bibinfo
  {pages} {2733} (\bibinfo {year} {1998})}\BibitemShut {NoStop}%
\bibitem [{\citenamefont {Kielpinski}\ \emph {et~al.}(2001)\citenamefont
  {Kielpinski}, \citenamefont {Meyer}, \citenamefont {Rowe}, \citenamefont
  {Sackett}, \citenamefont {Itano}, \citenamefont {Monroe},\ and\ \citenamefont
  {Wineland}}]{wineland_memory}%
  \BibitemOpen
  \bibfield  {author} {\bibinfo {author} {\bibfnamefont {D.}~\bibnamefont
  {Kielpinski}}, \bibinfo {author} {\bibfnamefont {V.}~\bibnamefont {Meyer}},
  \bibinfo {author} {\bibfnamefont {M.~A.}\ \bibnamefont {Rowe}}, \bibinfo
  {author} {\bibfnamefont {C.~A.}\ \bibnamefont {Sackett}}, \bibinfo {author}
  {\bibfnamefont {W.~M.}\ \bibnamefont {Itano}}, \bibinfo {author}
  {\bibfnamefont {C.}~\bibnamefont {Monroe}}, \ and\ \bibinfo {author}
  {\bibfnamefont {D.~J.}\ \bibnamefont {Wineland}},\ }\href {\doibase
  10.1126/science.1057357} {\bibfield  {journal} {\bibinfo  {journal}
  {Science}\ }\textbf {\bibinfo {volume} {291}},\ \bibinfo {pages} {1013}
  (\bibinfo {year} {2001})}\BibitemShut {NoStop}%
\bibitem [{\citenamefont {Viola}\ \emph {et~al.}(2001)\citenamefont {Viola},
  \citenamefont {Fortunato}, \citenamefont {Pravia}, \citenamefont {Knill},
  \citenamefont {Laflamme},\ and\ \citenamefont {Cory}}]{viola_noiseless}%
  \BibitemOpen
  \bibfield  {author} {\bibinfo {author} {\bibfnamefont {L.}~\bibnamefont
  {Viola}}, \bibinfo {author} {\bibfnamefont {E.~M.}\ \bibnamefont
  {Fortunato}}, \bibinfo {author} {\bibfnamefont {M.~A.}\ \bibnamefont
  {Pravia}}, \bibinfo {author} {\bibfnamefont {E.}~\bibnamefont {Knill}},
  \bibinfo {author} {\bibfnamefont {R.}~\bibnamefont {Laflamme}}, \ and\
  \bibinfo {author} {\bibfnamefont {D.~G.}\ \bibnamefont {Cory}},\ }\href
  {\doibase 10.1126/science.1064460} {\bibfield  {journal} {\bibinfo  {journal}
  {Science}\ }\textbf {\bibinfo {volume} {293}},\ \bibinfo {pages} {2059}
  (\bibinfo {year} {2001})}\BibitemShut {NoStop}%
\bibitem [{\citenamefont {Vitali}\ \emph {et~al.}(1997)\citenamefont {Vitali},
  \citenamefont {Tombesi},\ and\ \citenamefont {Milburn}}]{milburn_feedback}%
  \BibitemOpen
  \bibfield  {author} {\bibinfo {author} {\bibfnamefont {D.}~\bibnamefont
  {Vitali}}, \bibinfo {author} {\bibfnamefont {P.}~\bibnamefont {Tombesi}}, \
  and\ \bibinfo {author} {\bibfnamefont {G.~J.}\ \bibnamefont {Milburn}},\
  }\href {\doibase 10.1103/PhysRevLett.79.2442} {\bibfield  {journal} {\bibinfo
   {journal} {Phys. Rev. Lett.}\ }\textbf {\bibinfo {volume} {79}},\ \bibinfo
  {pages} {2442} (\bibinfo {year} {1997})}\BibitemShut {NoStop}%
\bibitem [{\citenamefont {Shulman}\ \emph {et~al.}(2014)\citenamefont
  {Shulman}, \citenamefont {Harvey}, \citenamefont {Nichol}, \citenamefont
  {Bartlett}, \citenamefont {Doherty}, \citenamefont {Umansky},\ and\
  \citenamefont {Yacoby}}]{control_2}%
  \BibitemOpen
  \bibfield  {author} {\bibinfo {author} {\bibfnamefont {M.~D.}\ \bibnamefont
  {Shulman}}, \bibinfo {author} {\bibfnamefont {S.~P.}\ \bibnamefont {Harvey}},
  \bibinfo {author} {\bibfnamefont {J.~M.}\ \bibnamefont {Nichol}}, \bibinfo
  {author} {\bibfnamefont {S.~D.}\ \bibnamefont {Bartlett}}, \bibinfo {author}
  {\bibfnamefont {A.~C.}\ \bibnamefont {Doherty}}, \bibinfo {author}
  {\bibfnamefont {V.}~\bibnamefont {Umansky}}, \ and\ \bibinfo {author}
  {\bibfnamefont {A.}~\bibnamefont {Yacoby}},\ }\href {\doibase
  10.1038/ncomms6156} {\bibfield  {journal} {\bibinfo  {journal} {Nature
  Communications}\ }\textbf {\bibinfo {volume} {5}},\ \bibinfo {pages} {5156}
  (\bibinfo {year} {2014})}\BibitemShut {NoStop}%
\bibitem [{\citenamefont {Ma\^{\i}tre}\ \emph {et~al.}(1997)\citenamefont
  {Ma\^{\i}tre}, \citenamefont {Hagley}, \citenamefont {Nogues}, \citenamefont
  {Wunderlich}, \citenamefont {Goy}, \citenamefont {Brune}, \citenamefont
  {Raimond},\ and\ \citenamefont {Haroche}}]{haroche_memory}%
  \BibitemOpen
  \bibfield  {author} {\bibinfo {author} {\bibfnamefont {X.}~\bibnamefont
  {Ma\^{\i}tre}}, \bibinfo {author} {\bibfnamefont {E.}~\bibnamefont {Hagley}},
  \bibinfo {author} {\bibfnamefont {G.}~\bibnamefont {Nogues}}, \bibinfo
  {author} {\bibfnamefont {C.}~\bibnamefont {Wunderlich}}, \bibinfo {author}
  {\bibfnamefont {P.}~\bibnamefont {Goy}}, \bibinfo {author} {\bibfnamefont
  {M.}~\bibnamefont {Brune}}, \bibinfo {author} {\bibfnamefont {J.~M.}\
  \bibnamefont {Raimond}}, \ and\ \bibinfo {author} {\bibfnamefont
  {S.}~\bibnamefont {Haroche}},\ }\href {\doibase 10.1103/PhysRevLett.79.769}
  {\bibfield  {journal} {\bibinfo  {journal} {Phys. Rev. Lett.}\ }\textbf
  {\bibinfo {volume} {79}},\ \bibinfo {pages} {769} (\bibinfo {year}
  {1997})}\BibitemShut {NoStop}%
\bibitem [{\citenamefont {Kraus}\ \emph {et~al.}(2006)\citenamefont {Kraus},
  \citenamefont {Tittel}, \citenamefont {Gisin}, \citenamefont {Nilsson},
  \citenamefont {Kr\"oll},\ and\ \citenamefont {Cirac}}]{reversible_1}%
  \BibitemOpen
  \bibfield  {author} {\bibinfo {author} {\bibfnamefont {B.}~\bibnamefont
  {Kraus}}, \bibinfo {author} {\bibfnamefont {W.}~\bibnamefont {Tittel}},
  \bibinfo {author} {\bibfnamefont {N.}~\bibnamefont {Gisin}}, \bibinfo
  {author} {\bibfnamefont {M.}~\bibnamefont {Nilsson}}, \bibinfo {author}
  {\bibfnamefont {S.}~\bibnamefont {Kr\"oll}}, \ and\ \bibinfo {author}
  {\bibfnamefont {J.~I.}\ \bibnamefont {Cirac}},\ }\href {\doibase
  10.1103/PhysRevA.73.020302} {\bibfield  {journal} {\bibinfo  {journal} {Phys.
  Rev. A}\ }\textbf {\bibinfo {volume} {73}},\ \bibinfo {pages} {020302}
  (\bibinfo {year} {2006})}\BibitemShut {NoStop}%
\bibitem [{\citenamefont {Streltsov}\ \emph {et~al.}()\citenamefont
  {Streltsov}, \citenamefont {Adesso},\ and\ \citenamefont {Plenio}}]{coh}%
  \BibitemOpen
  \bibfield  {author} {\bibinfo {author} {\bibfnamefont {A.}~\bibnamefont
  {Streltsov}}, \bibinfo {author} {\bibfnamefont {G.}~\bibnamefont {Adesso}}, \
  and\ \bibinfo {author} {\bibfnamefont {M.~B.}\ \bibnamefont {Plenio}},\
  }\href {https://arxiv.org/abs/1609.02439} {\bibinfo  {journal}
  {arXiv:1609.02439}\ }\BibitemShut {NoStop}%
\bibitem [{Note1()}]{Note1}%
  \BibitemOpen
\bibfield  {journal} {  }\bibinfo {note} {We choose $\omega _j=\Delta _\omega
  j$ with $\Delta _\omega =\pi /1000$, and $j$ is an integer $j\in
  [-100000,100000]$}\BibitemShut {NoStop}%
\bibitem [{\citenamefont {Elsayed}\ and\ \citenamefont {Fine}(2013)}]{tarek}%
  \BibitemOpen
  \bibfield  {author} {\bibinfo {author} {\bibfnamefont {T.~A.}\ \bibnamefont
  {Elsayed}}\ and\ \bibinfo {author} {\bibfnamefont {B.~V.}\ \bibnamefont
  {Fine}},\ }\href {\doibase 10.1103/PhysRevLett.110.070404} {\bibfield
  {journal} {\bibinfo  {journal} {Phys. Rev. Lett.}\ }\textbf {\bibinfo
  {volume} {110}},\ \bibinfo {pages} {070404} (\bibinfo {year}
  {2013})}\BibitemShut {NoStop}%
\bibitem [{\citenamefont {Steinigeweg}\ \emph {et~al.}(2014)\citenamefont
  {Steinigeweg}, \citenamefont {Khodja}, \citenamefont {Niemeyer},
  \citenamefont {Gogolin},\ and\ \citenamefont {Gemmer}}]{steinigeweg}%
  \BibitemOpen
  \bibfield  {author} {\bibinfo {author} {\bibfnamefont {R.}~\bibnamefont
  {Steinigeweg}}, \bibinfo {author} {\bibfnamefont {A.}~\bibnamefont {Khodja}},
  \bibinfo {author} {\bibfnamefont {H.}~\bibnamefont {Niemeyer}}, \bibinfo
  {author} {\bibfnamefont {C.}~\bibnamefont {Gogolin}}, \ and\ \bibinfo
  {author} {\bibfnamefont {J.}~\bibnamefont {Gemmer}},\ }\href {\doibase
  10.1103/PhysRevLett.112.130403} {\bibfield  {journal} {\bibinfo  {journal}
  {Phys. Rev. Lett.}\ }\textbf {\bibinfo {volume} {112}},\ \bibinfo {pages}
  {130403} (\bibinfo {year} {2014})}\BibitemShut {NoStop}%
\bibitem [{\citenamefont {Anderson}\ and\ \citenamefont
  {Weiss}(1953)}]{and_weiss}%
  \BibitemOpen
  \bibfield  {author} {\bibinfo {author} {\bibfnamefont {P.~W.}\ \bibnamefont
  {Anderson}}\ and\ \bibinfo {author} {\bibfnamefont {P.~R.}\ \bibnamefont
  {Weiss}},\ }\href {\doibase 10.1103/RevModPhys.25.269} {\bibfield  {journal}
  {\bibinfo  {journal} {Rev. Mod. Phys.}\ }\textbf {\bibinfo {volume} {25}},\
  \bibinfo {pages} {269} (\bibinfo {year} {1953})}\BibitemShut {NoStop}%
\bibitem [{\citenamefont {Braun}\ \emph {et~al.}(2013)\citenamefont {Braun},
  \citenamefont {Ronzheimer}, \citenamefont {Schreiber}, \citenamefont
  {Hodgman}, \citenamefont {Rom}, \citenamefont {Bloch},\ and\ \citenamefont
  {Schneider}}]{negative_temp}%
  \BibitemOpen
  \bibfield  {author} {\bibinfo {author} {\bibfnamefont {S.}~\bibnamefont
  {Braun}}, \bibinfo {author} {\bibfnamefont {J.~P.}\ \bibnamefont
  {Ronzheimer}}, \bibinfo {author} {\bibfnamefont {M.}~\bibnamefont
  {Schreiber}}, \bibinfo {author} {\bibfnamefont {S.~S.}\ \bibnamefont
  {Hodgman}}, \bibinfo {author} {\bibfnamefont {T.}~\bibnamefont {Rom}},
  \bibinfo {author} {\bibfnamefont {I.}~\bibnamefont {Bloch}}, \ and\ \bibinfo
  {author} {\bibfnamefont {U.}~\bibnamefont {Schneider}},\ }\href {\doibase
  10.1126/science.1227831} {\bibfield  {journal} {\bibinfo  {journal}
  {Science}\ }\textbf {\bibinfo {volume} {339}},\ \bibinfo {pages} {52}
  (\bibinfo {year} {2013})}\BibitemShut {NoStop}%
\bibitem [{\citenamefont {Tarkhov}\ \emph {et~al.}()\citenamefont {Tarkhov},
  \citenamefont {Wimberger},\ and\ \citenamefont {Fine}}]{andrey}%
  \BibitemOpen
  \bibfield  {author} {\bibinfo {author} {\bibfnamefont {A.~E.}\ \bibnamefont
  {Tarkhov}}, \bibinfo {author} {\bibfnamefont {S.}~\bibnamefont {Wimberger}},
  \ and\ \bibinfo {author} {\bibfnamefont {B.~V.}\ \bibnamefont {Fine}},\
  }\href {https://arxiv.org/abs/1705.08176} {\bibinfo  {journal}
  {arXiv:1705.08176}\ }\BibitemShut {NoStop}%
\bibitem [{\citenamefont {Fine}\ \emph {et~al.}(2014)\citenamefont {Fine},
  \citenamefont {Elsayed}, \citenamefont {Kropf},\ and\ \citenamefont
  {de~Wijn}}]{absence}%
  \BibitemOpen
\bibfield  {journal} {  }\bibfield  {author} {\bibinfo {author} {\bibfnamefont
  {B.~V.}\ \bibnamefont {Fine}}, \bibinfo {author} {\bibfnamefont {T.~A.}\
  \bibnamefont {Elsayed}}, \bibinfo {author} {\bibfnamefont {C.~M.}\
  \bibnamefont {Kropf}}, \ and\ \bibinfo {author} {\bibfnamefont {A.~S.}\
  \bibnamefont {de~Wijn}},\ }\href {\doibase 10.1103/PhysRevE.89.012923}
  {\bibfield  {journal} {\bibinfo  {journal} {Phys. Rev. E}\ }\textbf {\bibinfo
  {volume} {89}},\ \bibinfo {pages} {012923} (\bibinfo {year}
  {2014})}\BibitemShut {NoStop}%
\end{thebibliography}
\end{document}